\documentclass[aps,prd,longbibliography,nofootinbib,notitlepage,reprint]{revtex4-2}
\pdfoutput=1
\usepackage[T1]{fontenc}
\usepackage{bm,amsmath,amssymb,xcolor,bbm,mathrsfs} 
\usepackage[pdftex]{graphicx}
\usepackage[colorlinks=true, pdfstartview=FitH, linkcolor=blue, citecolor=blue, urlcolor=blue]{hyperref}
\usepackage{slashed}

\DeclareMathOperator\diag{diag}
\DeclareMathOperator\sgn{sgn}
\DeclareMathOperator\Tr{Tr}
\DeclareMathOperator\Pf{Pf}
\DeclareMathOperator\rme{\mathrm{e}}

\newcommand{\der}{\partial}

\newcommand{\bep}{\begin{pmatrix}} 
\newcommand{\eep}{\end{pmatrix}}

\newcommand{\SU}{\text{SU}}
\newcommand{\SO}{\text{SO}}
\renewcommand{\O}{\text{O}}
\newcommand{\U}{\text{U}}
\newcommand{\1}{\mathbbm{1}}
\newcommand{\RR}{\mathbb{R}}
\newcommand{\CC}{\mathbb{C}}

\renewcommand{\epsilon}{\varepsilon}
\newcommand{\rmd}{\mathrm{d}}

\DeclareMathOperator\erf{erf}
\newcommand{\muu}{\widehat{\mu}}

\def\ba#1\ea{\begin{align}#1\end{align}}
\def\mkakko#1{\left( #1 \right)}
\def\ckakko#1{\left\{ #1 \right\}}
\def\kkakko#1{\left[ #1 \right]}
\def\wt#1{\widetilde{#1}}

\def\cc#1{#1}
\newcommand{\ccc}{}

\begin{document}
\title{Unitary matrix integral for \cc{QCD with real quarks} and the GOE-GUE crossover}

\author{Takuya Kanazawa}
\affiliation{Research and Development Group, Hitachi, Ltd., Kokubunji, Tokyo 185-8601, Japan}
\allowdisplaybreaks

\begin{abstract}
A unitary matrix integral that appears in the low-energy limit of \cc{QCD-like theories with quarks in real representations of the gauge group} at finite chemical potential is analytically evaluated and expressed as a pfaffian. Its application to the GOE-GUE crossover in random matrix theory is discussed. \cc{An analogous unitary integral for QCD-like theories with quarks in pseudoreal representations of the gauge group is also evaluated.}
\end{abstract}
\maketitle

\section{Introduction}
Matrix integrals appear in diverse fields of mathematics and theoretical physics (see \cite{Rossi:1996hs,Morozov:2009jv} for reviews). In particular they play important roles in random matrix theory (RMT) \cite{Guhr:1997ve,Mehtabook,Akemann:2011csh}. The list of most well-studied unitary matrix integrals includes, but is not limited to, the Brezin-Gross-Witten integral \cite{Brezin:1980rk,Gross:1980he} (also known as the Leutwyler-Smilga integral \cite{Leutwyler:1992yt}), the Harish-Chandra-Itzykson-Zuber integral \cite{Harish-Chandra:1957dhy,Itzykson:1979fi}, and the Berezin-Karpelevich integral \cite{Berezin1958,Guhr:1996vx,Jackson:1996jb}. They are relevant to quantum gravity, lattice gauge theory, Quantum Chromodynamics (QCD), quantum chaos and disordered mesoscopic systems. 

In QCD, due to spontaneous breaking of chiral symmetry, the low-energy physics may be described by a non-linear sigma model. In the so-called $\epsilon$-regime \cite{Gasser:1987ah,Leutwyler:1992yt}, exact zero modes of the Nambu-Goldstone modes dominate the partition function and the infinite-dimensional path integral reduces to a finite-dimensional integral over a coset space. It is known since olden times that there are three patterns of chiral symmetry breaking in QCD, depending on the representation of quarks \cite{Peskin:1980gc}. In QCD-like theories with quarks in \cc{real} representations of the gauge group, the pattern is $\SU(2N_f)\to\SO(2N_f)$ where $N_f$ is the number of Dirac fermions \cite{Smilga:1994tb,Kogut:2000ek}. The basic degrees of freedom at low energy are expressed through $U^T U$ where $U$ is a matrix field that lives on $\U(2N_f)$. When a quark chemical potential $\mu$ is added, the low-energy sigma model attains additional terms \cite{Kogut:2000ek}. In this paper, we are interesting in evaluating the $\epsilon$-regime partition function of \cc{QCD with real quarks} when the chemical potential is different for each flavor, i.e., $(\mu_1,\cdots,\mu_{N_f})$ are all distinct. This case was not covered in \cite{Kogut:2000ek}. We will show that the partition function has a pfaffian form. We also argue that the integral formula has an application to the symmetry crossover between the Gaussian Orthogonal Ensemble (GOE) and the Gaussian Unitary Ensemble (GUE) in RMT. Moreover, since $U^TU$ is an element of the Circular Orthogonal Ensemble in RMT \cite{Dyson:1962es}, our result may have a potential application in this direction as well. 

This paper is organized as follows. In section~\ref{sc:main} the main analytical results of this paper are summarized. In section~\ref{sc:appl} some applications are illustrated. In section~\ref{sc:deriv} we give a derivation of the integral formulae presented in section~\ref{sc:main}. \cc{In section~\ref{sc:added} we give a formula (without proof) for a related unitary matrix integral that has applications to QCD-like theories with quarks in pseudoreal representations of the gauge group.} Finally in section~\ref{sc:conc} we conclude. 

\section{\label{sc:main}Main results}

For an arbitrary Hermitian $N\times N$ matrix $H$ with mutually distinct eigenvalues $\{e_k\}$ and an arbitrary nonzero $\gamma\in\CC$, we have 
\ba
	& \int_{\U(N)} \!\!\!\! \rmd{U} \exp \kkakko{\gamma^2 \Tr(U^T U H U^\dagger U^* H^*)} 
	\notag
	\\
	= \; &  
	\kkakko{\prod_{k=1}^{N} \Gamma\mkakko{\frac{k+1}{2}}}
	\frac{\exp\mkakko{\displaystyle \gamma^2\sum_{k=1}^{N}e_k^2}}{\gamma^{N(N-1)/2}\Delta_N(e)}
	\notag
	\\
	& \times \underset{1\leq i,j\leq N}{\Pf}\Big[\erf\mkakko{\gamma(e_i-e_j)}\Big]
	\label{eq:main1}
\ea
for even $N$ and 
\ba
	& \int_{\U(N)} \!\!\!\! \rmd{U} \exp \kkakko{\gamma^2\Tr(U^T U H U^\dagger U^* H^*)} 
	\notag
	\\
	=\; & \kkakko{\prod_{k=1}^{N} \Gamma\mkakko{\frac{k+1}{2}}}
	\frac{\exp\mkakko{\displaystyle \gamma^2\sum_{k=1}^{N}e_k^2}}{\gamma^{N(N-1)/2}\Delta_N(e)}
	\notag
	\\
	& \times \Pf
	\kkakko{\begin{array}{c|c}
		\big\{\erf\mkakko{\gamma(e_i-e_j)}\big\}_{1\leq i,j\leq N} & \{1\}_{1 \leq i\leq N}
		\vspace{1pt}\\\hline 
		\{ -1 \}_{1\leq j \leq N} & 0
	\end{array}} 
	\label{eq:main2}
\ea
for odd $N$, where $\rmd U$ denotes the normalized Haar measure, $\displaystyle\Delta_N(e)\equiv \prod_{i<j}(e_i-e_j)$ is the Vandermonde determinant, $\Pf$ denotes a pfaffian, and $\erf(x)$ is the error function.

We performed an intensive numerical check of these formulae for $N$ up to $10$ by estimating the left hand sides of the formulae by Monte Carlo methods and verified their correctness. We used a Python library \textsf{pfapack} \cite{wimmer2012algorithm} for an efficient calculation of a pfaffian. 

As a side remark we note that a similar pfaffian formula has been obtained for a unitary matrix integral considered in \cite{Kanazawa:2018kbo}.

\section{\label{sc:appl}Applications}
\subsection{\label{sc:adjq}\cc{QCD with real quarks}}

It follows from a slight generalization of \cite{Kogut:2000ek} that the static part of the partition function for the low-energy limit of massless QCD with $N_f$ flavors \cc{of quarks in real representations} with chemical potential $\{\mu_f\}_{f=1}^{N_f}$ is given by
\ba
	Z = \int_{\U(2N_f)}\hspace{-6mm} \rmd U \exp \ckakko{V_4 F^2 \Tr[U^T U B (U^T U)^\dagger B + B^2]}
\ea
where $V_4$ is the Euclidean spacetime volume, $F$ is the pion decay constant, and 
\ba
	B\equiv \diag(\mu_1,\cdots,\mu_{N_f},-\mu_1,\cdots,-\mu_{N_f})
\ea 
is the chemical potential matrix. The same sigma model is expected to arise in the large-$N$ limit of the chiral symplectic Ginibre ensemble \cite{Akemann:2005fd} which has exactly the same symmetry as \cc{QCD with real quarks}. 

Let us define the dimensionless variables
\ba
	\left\{\begin{array}{rl}
		\muu_f & \equiv \sqrt{V_4} F \mu_f
		\\
		\muu_{N_f+f} & \equiv - \muu_f 
	\end{array}
	\quad \text{for}~f=1,2,\cdots,N_f\,. \right.
\ea
Then a straightforward application of \eqref{eq:main1} yields

\ba
	Z \propto  
	\frac{
		\exp\mkakko{\displaystyle 4\sum_{f=1}^{N_f}\muu_f^2}
		\underset{1\leq f,g\leq 2N_f}{\Pf}\big[\erf\mkakko{\muu_f - \muu_g}\big]
	}{
		\displaystyle
		\kkakko{\prod_{1\leq f<g\leq N_f}(\muu_f-\muu_g)^2(\muu_f+\muu_g)^2}
		\prod_{f=1}^{N_f}\muu_f
	}
\ea
which completely fixes the chemical potential dependence of the effective theory in the $\epsilon$-regime.

\subsection{\label{sc:0923234}GOE-GUE crossover}

In the classical papers \cite{Pandey:1982br,Mehta:1983ns}, Mehta and Pandey solved the random matrix ensemble intermediate between GOE and GUE. Their ingenious approach was to consider the random matrix
\ba
	H = S + \alpha T\,, \label{eq:MPrm}
\ea
where $S$ is a Gaussian real symmetric matrix and $T$ is a Gaussian Hermitian matrix. As $\alpha$ grows from zero, the level statistics evolve from GOE to GUE. The transition occurs at the scale $\alpha^2\sim 1/N$ where $N$ is the matrix size. Mehta and Pandey successfully derived the joint probability distribution function of eigenvalues of $H$ by using the Harish-Chandra-Itzykson-Zuber integral \cite{Harish-Chandra:1957dhy,Itzykson:1979fi}. 

An alternative approach to the GOE-GUE transition would be to consider the random matrix
\ba
	H = S + i \alpha A\,, \label{eq:werm}
\ea
where $A$ is a Gaussian real anti-symmetric matrix. [Actually the ensemble \eqref{eq:MPrm} is essentially equivalent to \eqref{eq:werm}, because the sum of two Gaussian real symmetric matrices is again a Gaussian real symmetric matrix with a modified variance.] Then the Gaussian weight for $S$ and $A$ reads as
\ba
	& \quad \exp(-\Tr S^2 + \Tr A^2) 
	\notag
	\\
	& = \exp\kkakko{
		-\frac{1+\alpha^2}{2\alpha^2}\Tr(H^2) + \frac{1-\alpha^2}{2\alpha^2} \Tr (HH^T)
	}.
\ea
Upon diagonalization $H=UEU^\dagger$ we end up with the unitary integral which exactly has the form \eqref{eq:main1} and \eqref{eq:main2}. Carrying out the integrals, we immediately arrive at the joint eigenvalue density derived by Mehta and Pandey \cite{Pandey:1982br,Mehta:1983ns}. Our integral thus provides a way to solve the transitive ensemble without recourse to the Harish-Chandra-Itzykson-Zuber integral.

\section{\label{sc:deriv}Derivation of the formulae}

The derivation proceeds in three steps.

\subsection{Step 1:~the heat equation}

We adopt the method of heat equation \cite{Itzykson:1979fi}. Let us assume $t>0$ and consider a function
\ba
	z_N(t,H,U) & \equiv \frac{1}{t^{\alpha}} 	\exp\kkakko{-\frac{1}{t}\Tr(H-P^\dagger H^TP)^2}
\ea
where we wrote $P=U^T U$ for brevity. Then
\ba
	\frac{\der}{\der t}z_N(t,H,U) & = \kkakko{-\frac{\alpha}{t} + \frac{1}{t^2}\Tr(H-P^\dagger H^TP)^2}
	\notag
	\\
	& \quad \times z_N(t,H,U)\,.
	\label{eq:zdert}
\ea
The Laplacian over Hermitian matrices is defined by 
\ba
	\Delta_H & \equiv \sum_{i}\frac{\der^2}{\der H_{ii}^2} + \frac{1}{2}
	\sum_{i<j}\kkakko{\frac{\der^2}{\der(\mathrm{Re}\,H_{ij})^2}+\frac{\der^2}{\der(\mathrm{Im}\,H_{ij})^2}}
	\hspace{-2mm}
	\\
	& = \sum_{i,j}\frac{\der}{\der H_{ij}}\frac{\der}{\der H_{ji}}\,.
\ea
Then we have (assuming that repeated indices are summed)
\ba
	& \quad \Delta_H z_N(t,H,U) 
	\notag
	\\
	& = \frac{1}{t^\alpha}\frac{\der}{\der H_{ij}}\frac{\der}{\der H_{ji}}
	\exp\kkakko{-\frac{1}{t}\Tr(H-P^\dagger H^TP)^2}
	\\
	& = \kkakko{-\frac{4N(N-1)}{t} + \frac{16}{t^2}\Tr(H-P^\dagger H^T P)^2 } z_N(t,H,U)\,.
	\label{eq:zderH}
\ea
Comparison of \eqref{eq:zdert} and \eqref{eq:zderH} indicates that, if we set $\displaystyle \alpha=N(N-1)/4$, then
\ba
	\mkakko{\frac{\der}{\der t} - \frac{1}{16}\Delta_H} z_N(t,H,U) = 0
\ea
holds. Then it is obvious that
\ba
	Z_N(t,H) & \equiv \int_{\U(N)} \!\!\!\! \rmd{U}~ z_N(t,H,U)
	\\
	& = \frac{1}{t^\alpha}\exp\mkakko{-\frac{2}{t}\Tr H^2} 
	\notag
	\\
	& \quad \times 
	\int_{\U(N)} \!\!\!\! \rmd{U} \exp \kkakko{\frac{2}{t}\Tr(PHP^\dagger H^T)}
	\label{eq:ZNdefu}
\ea
also satisfies the same differential equation as $z_N(t,H,U)$. Using the invariance of the Haar measure it is easy to verify that $Z_N(t,H)$ depends on $H$ only through its eigenvalues $\{e_1,\cdots,e_N\}$. An important property of $Z_N(t,H)$ is its translational invariance. Namely, $Z_N(t,H)=Z_N(t,H+a\1_N)$ for an arbitrary $a\in\RR$, which can be easily verified from the definition \eqref{eq:ZNdefu}. This means that $Z_N(t,H)$ depends on $\{e_k\}$ only through the differences $\{e_k-e_\ell \}$. 

For us it is beneficial to transform the Laplacian into the ``polar coordinate'' \cite{ZinnJustin:2002pk} 
\ba
	\Delta_H = \frac{1}{\Delta_N(e)}\sum_{k=1}^{N}\frac{\der^2}{\der e_k^2}\Delta_N(e) + \Delta_X
\ea
where $X$ denote the angular variables. Then
\ba
	\mkakko{\frac{\der}{\der t}-\frac{1}{16}\sum_{k=1}^{N}\frac{\der^2}{\der e_k^2}}[\Delta_N(e)Z_N(t,H)] = 0\,.
	\label{eq:differ3242}
\ea
To obtain the basic building block of $Z_N(t,H)$, we shall explicitly work out the $N=2$ case in the next subsection.

\subsection{\boldmath Step 2:~the $N=2$ case}

Let us look at $N=2$, for which
\ba
	Z_2(t,H) & = \frac{1}{\sqrt{t}}\exp\kkakko{-\frac{2}{t}(e_1^2+e_2^2)}
	\notag
	\\
	& \quad \times 
	\int_{\SU(2)} \!\!\!\! \rmd{U} \exp\ckakko{\frac{2}{t}\Tr[(U^TU)E(U^TU)^\dagger E]}
\ea
where $E=\diag(e_1,e_2)$ and we dropped the $\U(1)$ phase of $U$ because it decouples. Using the parametrization $U=x_0\1_2 + i x_k \sigma_k$ with $x_0^2+x_1^2+x_2^2+x_3^2=1$, it is easily verified that 
\ba
	& \quad \Tr[(U^TU)E(U^TU)^\dagger E] 
	\notag \\
	& = e_1^2+e_2^2-4(e_1-e_2)^2 (x_0x_1+x_2x_3)^2,
\ea
hence
\ba
	Z_2(t,H) = \frac{1}{\sqrt{t}}
	\int_{\SU(2)} \!\!\!\! \rmd{U} \exp\ckakko{-\frac{8}{t}(e_1-e_2)^2 (x_0x_1+x_2x_3)^2}\,.
\ea
Next we employ the Hopf coordinate of $S^3$:
\ba
	& (x_0,x_1,x_2,x_3) 
	\notag
	\\
	= \; & (\cos \xi_1\sin \eta, \cos \xi_2 \cos\eta, \sin \xi_1 \sin \eta, \sin \xi_2 \cos\eta)
\ea
which yields
\ba
	Z_2(t,H) & = \frac{1}{2\pi^2\sqrt{t}}
	\int_0^{2\pi}\rmd \xi_1 \int_0^{2\pi}\rmd \xi_2 \int_0^{\frac{\pi}{2}}\rmd \eta ~ \sin \eta \cos\eta  
	\notag
	\\
	& \quad \times \exp\ckakko{-\frac{2}{t}(e_1-e_2)^2\sin^{2}2\eta \cos^2(\xi_1-\xi_2)}
	\\
	& = \frac{1}{4\pi\sqrt{t}}
	\int_{-1}^{1}\rmd (\cos 2\eta)  \int_0^{2\pi}\rmd \xi 
	\notag
	\\
	& \quad \times 
	\exp\ckakko{-\frac{2}{t}(e_1-e_2)^2\sin^{2}2\eta \cos^{2}\xi}
	\\
	& = \frac{1}{2\sqrt{t}}
	\int_{-1}^{1}\rmd s~ \exp\kkakko{-\frac{1-s^2}{t}(e_1-e_2)^2}
	\notag
	\\
	& \quad \times 
	I_0 \mkakko{\frac{1-s^2}{t}(e_1-e_2)^2}
	\\
	& = \frac{1}{2\sqrt{t}}D\mkakko{\frac{2}{t}(e_1-e_2)^2}
	\\
	& =: Z_2(t,e_1,e_2)\,,
	\label{eq:254r232}
\ea
where we have defined 
\ba
	D(x) & \equiv \int_{-1}^{1}\rmd s~\exp\mkakko{-\frac{1-s^2}{2}x}I_0\mkakko{\frac{1-s^2}{2}x}.
\ea 
From \eqref{eq:differ3242}, we know that \eqref{eq:254r232} fulfills the equation
\ba
	\kkakko{\frac{\der}{\der t}-\frac{1}{16}\mkakko{\frac{\der^2}{\der e_1^2}+\frac{\der^2}{\der e_2^2}}}(e_1-e_2)Z_2(t,e_1,e_2) = 0\,.
\ea
It is then easy to find a solution for $N=4$ by treating $Z_2$ as a building block, e.g.,
\begin{multline}
	\kkakko{\frac{\der}{\der t}-\frac{1}{16}
	\sum_{k=1}^{4}\frac{\der^2}{\der e_k^2}}
	(e_1-e_2)Z_2(t,e_1,e_2)
	\\
	\times (e_3-e_4)Z_2(t,e_3,e_4) = 0 
\end{multline}
and so on. However, the solution $\Delta_N(e)Z_N(t,H)$ that we want should be antisymmetric under the exchange of any two $e_k$'s, so we are naturally led to the Pfaffian 
\ba
	\Delta_N(e)Z_N(t,H) & = \wt{C}_N \underset{1\leq i,j\leq N}{\Pf}
	\kkakko{(e_i-e_j)Z_2(t,e_i,e_j)} \!\!\!\!
	\label{eq:6534234}
\ea
for even $N$ and 
\begin{multline}
	\Delta_N(e)Z_N(t,H) = 
	\\
	\wt{C}_N \Pf
	\kkakko{
		\begin{array}{c|c}
			\Big\{(e_i-e_j)Z_2(t,e_i,e_j)\Big\}_{1\leq i,j\leq N} 
			& \{1\}_{1\leq i \leq N}
			\vspace{1pt}
			\\\hline
			\{ -1 \}_{1\leq j \leq N} & 0
		\end{array}
	} \!\!\!\!\!\!
	\label{eq:192475}
\end{multline}
for odd $N$, where $\wt{C}_N$ is the not yet determined constant. By construction, $\wt{C}_2=1$. We note that this Pfaffian form is consistent with the translational invariance of $Z_N(t,H)$. To fix the normalization uniquely, we need to examine the boundary condition at $t=+0$.

\subsection{\boldmath Step 3:~saddle point analysis}

To determine the normalization of the Pfaffian formula, let us perform the saddle point approximation of
\begin{multline}
	Z_N(t,H) = \frac{1}{t^\alpha}\exp\mkakko{-\frac{2}{t}\Tr H^2} 
	\\
	\times 
	\int_{\U(N)} \!\!\!\! \rmd{U} \exp \kkakko{\frac{2}{t}\Tr(U^T U E U^\dagger U^* E)} 
	\label{eq:4322dsf}
\end{multline}
for $t\to +0$, with $E=\diag(e_1,e_2,\cdots,e_N)$. Apparently $U=\1_N$ is the dominant saddle point. A closer inspection of the exponent of \eqref{eq:4322dsf} shows that, in fact, any $U$ of the form $OV$ with $O\in \O(N)$ and $V=\diag(\rme^{i\theta_1},\cdots,\rme^{i\theta_N})\in\U(1)^N$ is also a saddle. Hence the saddle point manifold appears to be given by $\U(1)^N\times \O(N)$. However there is a subtlety: the $2^N$ elements $\diag(\pm 1,\cdots,\pm 1)$ belong to both $\U(1)^N$ and $\O(N)$. To avoid duplication, the correct saddle point manifold must be considered as $[\U(1)^N \times \O(N)]/\{\pm 1\}^N$. When performing the saddle point integral, the volume of this manifold should be factored out, because the Gaussian integral is performed only for massive modes. Now the part of $\U(N)$ over which the Gaussian integration is done may be parametrized as $U=\exp(i\sum_{i<j}\theta_{ij}S_{ij})$, where $S_{ij}$ is a symmetric matrix such that $(S_{ij})_{pq}=\delta_{ip}\delta_{jq}+\delta_{iq}\delta_{jp}$. Then
\ba
	& \Tr(U^T U E U^\dagger U^* E) 
	\notag 
	\\
	= \; & \Tr \kkakko{E^2+4(\theta_{ij}S_{ij}E)^2 - 4 (\theta_{ij}S_{ij})^2 E^2}
	+ O(\theta^3)
	\\
	= \; & \sum_{k}e_k^2 - 4 \sum_{k<\ell}(e_k-e_\ell)^2 \theta_{k\ell}^2
	+ O(\theta^3)\,.
\ea
Therefore
\ba
	Z_N(t,H) & \approx \frac{1}{t^\alpha} 
	\frac{\mathrm{Vol}([\U(1)^N \times \O(N)]/\{\pm 1\}^N)}{\mathrm{Vol}(\U(N))}
	\notag
	\\
	& \quad \times 
	\int_{-\infty}^{\infty} \prod_{i<j}\rmd \theta_{ij} \exp \mkakko{
		-\frac{8}{t}\sum_{k<\ell}(e_k-e_\ell)^2 \theta_{k\ell}^2
	}
	\\
	& = \frac{1}{t^\alpha} 
	\frac{\mathrm{Vol}([\U(1)^N \times \O(N)]/\{\pm 1\}^N)}{\mathrm{Vol}(\U(N))}
	\notag
	\\
	& \quad \times 
	\mkakko{\frac{\pi t}{8}}^{N(N-1)/4}\frac{1}{|\Delta_N(e)|}\,.
\ea
Recalling $\alpha=N(N-1)/4$ and using the volume formulae \cite{Boya:2003km,Zyczkowski_2003,zhang2015volumes}
\ba
	\mathrm{Vol}(\U(N)) = \frac{2^N \pi^{N(N+1)/2}}{\prod_{k=1}^{N}\Gamma(k)}\,, ~
	\mathrm{Vol}(\O(N)) = \frac{2^N \pi^{N(N+1)/4}}{\prod_{k=1}^{N}\Gamma\mkakko{\frac{k}{2}}}
\ea
as well as the duplication formula $\displaystyle \Gamma\left(2z\right)=\pi^{-1/2}2^{2z-1}\Gamma\left(z\right)\Gamma\left(z+\tfrac{1}{2}\right)$ \cite{DLMF_gamma}, we finally obtain
\ba
	Z_N(t,H) & \approx 2^{-N(N-1)/4}\kkakko{\prod_{k=1}^{N} \Gamma\mkakko{\frac{k+1}{2}}}
	\frac{1}{|\Delta_N(e)|}\,. \!\!\!
	\label{eq:zlimit}
\ea
Note that this formula is valid for both even $N$ and odd $N$. 

Next we turn to \eqref{eq:6534234} for even $N$. As $D(x)\sim \sqrt{\pi/x}$ for $x\gg 1$, $Z_2(t,e_1,e_2)\sim \frac{\sqrt{\pi}}{2\sqrt{2}}\frac{1}{|e_1-e_2|}$. Substituting this into \eqref{eq:6534234} yields 
\ba
	Z_N(t,H) & \approx 2^{-3N/4}\pi^{N/4}
	\frac{\wt{C}_N}{\Delta_N(e)} \underset{1\leq i,j\leq N}{\Pf}
	\kkakko{\sgn(e_i-e_j)}
	\\
	& = 2^{-3N/4}\pi^{N/4}
	\frac{\wt{C}_N}{|\Delta_N(e)|}\,.
	\label{eq:90853432}
\ea
Matching \eqref{eq:zlimit} and \eqref{eq:90853432} we obtain
\ba
	\wt{C}_N & = 2^{-N(N-4)/4}\pi^{-N/4}
	\prod_{k=1}^{N} \Gamma\mkakko{\frac{k+1}{2}}
\ea
for even $N$. Similarly, for odd $N$, \eqref{eq:192475} yields 
\ba
	Z_N(t,H) & \approx 2^{-3(N-1)/4} \pi^{(N-1)/4} \frac{\wt{C}_N}{|\Delta_N(e)|}\,.
\ea
which, combined with \eqref{eq:zlimit}, leads to
\ba
	\wt{C}_N & = 2^{-(N-1)(N-3)/4} \pi^{-(N-1)/4} \prod_{k=1}^{N} \Gamma\mkakko{\frac{k+1}{2}}
\ea
for odd $N$. 

Finally, introducing $\gamma\equiv \sqrt{2/t}$ and using the formula 
\ba
	\frac{x}{\sqrt{\pi}}D(x^2) = \erf(x) \,,
	\label{eq:254wer}
\ea
we find that \eqref{eq:6534234} and \eqref{eq:192475} are equivalent to \eqref{eq:main1} and \eqref{eq:main2}, respectively. Thus we have proved \eqref{eq:main1} and \eqref{eq:main2} for $\gamma>0$. Since both sides of the equations \eqref{eq:main1} and \eqref{eq:main2} are analytic in $\gamma$, it is concluded that these equations are valid also for complex $\gamma$. This completes the proof.

The equality \eqref{eq:254wer} does not seem to be known in the literature, but can be shown by noting that both sides of \eqref{eq:254wer} solve the heat equation with the same initial condition.

{\ccc 
\section{\label{sc:added}\mbox{\cc{Related integral}}}

The methodology employed to prove \eqref{eq:main1}  and \eqref{eq:main2} can be put to work to prove another integral formula with no fundamental difficulty. Let us state the main result. The proof has been published in \cite{Kanazawa:2021oej}. 

Let $N$ be an even positive integer and define an $N\times N$ antisymmetric matrix
\ba
	I \equiv \diag\mkakko{
		\begin{pmatrix}0&1\\-1& 0\end{pmatrix}, 
		\cdots, 
		\begin{pmatrix}0&1\\-1& 0\end{pmatrix}
	}\,.
	\label{eq:Idefine}
\ea
Then, for an arbitrary Hermitian $N\times N$ matrix $H$ with mutually distinct eigenvalues $\{e_k\}$ and an arbitrary nonzero $\beta\in\CC$, we have 
\ba
	& \int_{\U(N)} \!\!\!\! \rmd{U} \exp \kkakko{\beta \Tr(U^T I U H (U^T I U)^\dagger H^*)} 
	\notag
	\\
	= \; &  
	\frac{\prod_{k=1}^{N/2} \Gamma\mkakko{2k-1}}{(2\beta)^{N(N-2)/4}\Delta_N(e)}
	\underset{1\leq i,j\leq N}{\Pf}\big[
	(e_i-e_j) \rme^{2\beta e_i e_j}
	\big]\,,
	\label{eq:77777}
\ea
where $\rmd U$ denotes the normalized Haar measure, $\Pf$ denotes a pfaffian, and $\displaystyle\Delta_N(e)\equiv \prod_{i<j}(e_i-e_j)$ is the Vandermonde determinant. We performed an intensive numerical check of \eqref{eq:77777} for $N\in\{2,4,6,8,10\}$ with Monte Carlo methods and confirmed its correctness. Equation \eqref{eq:77777} enables us to compute the $\varepsilon$-regime partition function of massless QCD-like theories with pseudoreal quarks at finite chemical potential. Moreover, \eqref{eq:77777} simplifies the study of the GSE-GUE crossover in random matrix theory, as in section~\ref{sc:0923234}.
}

\section{\label{sc:conc}Conclusions and outlook}
In this paper we have derived a pfaffian formula for a unitary matrix integral with the heat equation method, and used it to evaluate the low-energy partition function of \cc{QCD-like theories with reak quarks}. We also showed that the classical result by Mehta and Pandey for the transitive ensemble between GOE and GUE can be reproduced with our formula. There are miscellaneous future directions of research, which we list below:
\begin{itemize}
	\item[$\checkmark$] Our integral is a special case of a more general integral 
	\ba
		\int_{\U(N)}\!\!\!\!\rmd U \exp\kkakko{\Tr(U^T S U H U^\dagger S U^* H^*)}
	\ea
	where $H$ is a Hermitian matrix and $S$ is a real symmetric matrix. Putting $S=\1_N$ reproduces our integral. It is easily seen that this integral is a function of the eigenvalues of $S$ and $H$ only. It would be interesting to seek for an analytic formula for this integral. 
	\item[$\checkmark$] In section~\ref{sc:adjq} we considered the partition function in the massless limit. Whether our integral formulae can be extended to the massive case is an open problem.
	\item[$\checkmark$] An extension to supergroups along the lines of \cite{Guhr1991,Alfaro:1994ca,Guhr:1996mx} would be interesting. 
	\item[$\checkmark$] Our formula is valid only for Hermitian $H$. Can we relax this constraint and generalize the formula to an arbitrary complex matrix? Methods such as those in \cite{Guhr:1996vx,Balantekin:2000vn} may prove helpful in this regard. 
\end{itemize}

\acknowledgments
We thank J.J.M.~Verbaarschot and M.~Kieburg for helpful comments.

\bibliography{draft_new_integral_v3.bbl}
\end{document}